\documentclass[aps,prl,twocolumn,showpacs]{revtex4-1}
\usepackage[utf8]{inputenc}
\usepackage{graphicx}
\usepackage{amsmath}
\usepackage{hyperref}
\usepackage{units}
\usepackage{amsmath, amssymb}
\usepackage{natbib}
\usepackage{amsfonts}
\usepackage{textcomp}
\usepackage{verbatim}
\usepackage{natbib}
\usepackage{color}

\begin{document}

\title{Softening of Roton and Phonon Modes  in a Bose-Einstein Condensate with Spin-Orbit Coupling}
\author{Si-Cong Ji$^{\star\star}$}
\affiliation{Hefei National Laboratory for Physical Sciences at Microscale and Department of Modern Physics, Synergetic Innovation Center of Quantum Information and Quantum Physics, 
 University of Science and Technology of China, Hefei, Anhui 230026, China}
\author{Long Zhang$^{\star\star}$}
\affiliation{Hefei National Laboratory for Physical Sciences at Microscale and Department of Modern Physics, Synergetic Innovation Center of Quantum Information and Quantum Physics, 
 University of Science and Technology of China, Hefei, Anhui 230026, China}
\author{Xiao-Tian Xu}
\affiliation{Hefei National Laboratory for Physical Sciences at Microscale and Department of Modern Physics, Synergetic Innovation Center of Quantum Information and Quantum Physics, 
 University of Science and Technology of China, Hefei, Anhui 230026, China} 
 \author{Zhan Wu}
\affiliation{Hefei National Laboratory for Physical Sciences at Microscale and Department of Modern Physics, Synergetic Innovation Center of Quantum Information and Quantum Physics, 
 University of Science and Technology of China, Hefei, Anhui 230026, China} 
 \author{Youjin Deng}
\affiliation{Hefei National Laboratory for Physical Sciences at Microscale and Department of Modern Physics, Synergetic Innovation Center of Quantum Information and Quantum Physics, 
 University of Science and Technology of China, Hefei, Anhui 230026, China} 
\author{Shuai Chen}
\email{shuai@ustc.edu.cn}
\affiliation{Hefei National Laboratory for Physical Sciences at Microscale and Department of Modern Physics, Synergetic Innovation Center of Quantum Information and Quantum Physics, 
 University of Science and Technology of China, Hefei, Anhui 230026, China} 
\author{Jian-Wei Pan}
\email{pan@ustc.edu.cn}
\affiliation{Hefei National Laboratory for Physical Sciences at Microscale and Department of Modern Physics, Synergetic Innovation Center of Quantum Information and Quantum Physics, 
 University of Science and Technology of China, Hefei, Anhui 230026, China} 
\begin{abstract}
Roton-type excitations usually emerge from strong correlations or long-range interactions, as in superfluid helium or dipolar ultracold atoms. 
However, in weakly short-range interacting quantum gas, the recently synthesized spin-orbit (SO) coupling can lead to 
various unconventional phases of superfluidity, and give rise to an excitation spectrum of roton-maxon character. 
Using Bragg spectroscopy we study a SO coupled Bose-Einstein condensate of $^{87}$Rb atoms,
and show that the excitation spectrum in a ``magnetized" phase clearly possesses a two-branch and roton-maxon structure.
As Raman coupling strength $\Omega$ is decreased, a roton-mode softening is observed, as a precursor of the phase transition to a stripe phase 
that spontaneously breaks spatially translational symmetry.
The measured roton gaps agree well with theoretical calculations. 
Further, we determine sound velocities both in the magnetized and the non-magnetized phase, 
and a phonon-mode softening is observed around the phase transition in between.
The validity of the $f$-sum rule is examined.
\end{abstract}
\maketitle

Roton and phonon are two typical excitation modes of superfluids.
They were first introduced by Landau in his phenomenological explanation on superfluidity of liquid helium~\cite{Landau}, 
and an experimental observation was realized  about two decades later~\cite{helium}. 
The emergence of roton mode in superfluid helium originates from strong density correlations. 
In weakly interacting ultrocold gases, long-range dipole-dipole interactions
can induce a roton-maxson dispersion~\cite{ODell2003,Shlyapnikov2003},
which were recently observed in a Bose-Einstein condensate (BEC) with cavity-mediated long-range interactions~\cite{Esslinger2012}. 
Across the phase transition from a superfluid to a supersolid phase, a softening of roton mode was further demonstrated~\cite{Esslinger2012}.
An important question naturally arises: can an excitation spectrum of roton-maxon character 
be observed in a quantum gas with weak and short-range interactions? 

Recently, artificial one-dimensional SO coupling has been synthesized in ultracold bosonic~\cite{Spielman2011,Jinyi2012} and fermonic~\cite{Jing2012,Zwierlein2012} atoms 
by two counter-propagating Raman lasers that couple the momentum of an atom to its spin~\cite{Spielman2013}. 
The single-particle dispersion is significantly modified such that  
a degenerate double-well structure appears for some Raman-coupling strength $\Omega$. 
Despite that interatomic interactions are weak and short-ranged, 
these systems can exhibit many unconventional condensate phases.
For $^{87}$Rb atoms, as $\Omega$ increases, the ground-state phase diagram is predicted to include~\cite{HoZhang2011,Yun2012}: 
a stripe phase of periodic density fringes that breaks translational symmetry, 
a ``magnetized" phase breaking a discrete $Z_2$ symmetry, and a non-magnetic phase. 
This rich structure of phase diagram has been largely supported by experiments~\cite{Spielman2011}, 
and its finite-temperature analog has also been explored~\cite{Cong2014}.

These recent vast experimental progresses in manipulating spin-orbit (SO) coupling pave a way to address the aforementioned question.
It is recognized that superfluids with a tendency towards periodic order can have phonon- and roton-type excitation modes.
In SO-coupled condensate of $^{87}$Rb atoms, the occurrence of the stripe phase preceding the magnetized phase
indicates that the excitation spectrum in the latter exhibits a roton-maxon structure~\cite{Stringari2012,zhengwei}. 
As the phase boundary is approached,  a roton-mode softening is further expected.
In this work, we experimentally demonstrate such a structure, and thus 
provide the first direct experimental observation of roton mode and its softening in weak and short-range interacting systems.
 
\begin{figure}
\centering
\includegraphics[width=0.90\columnwidth]{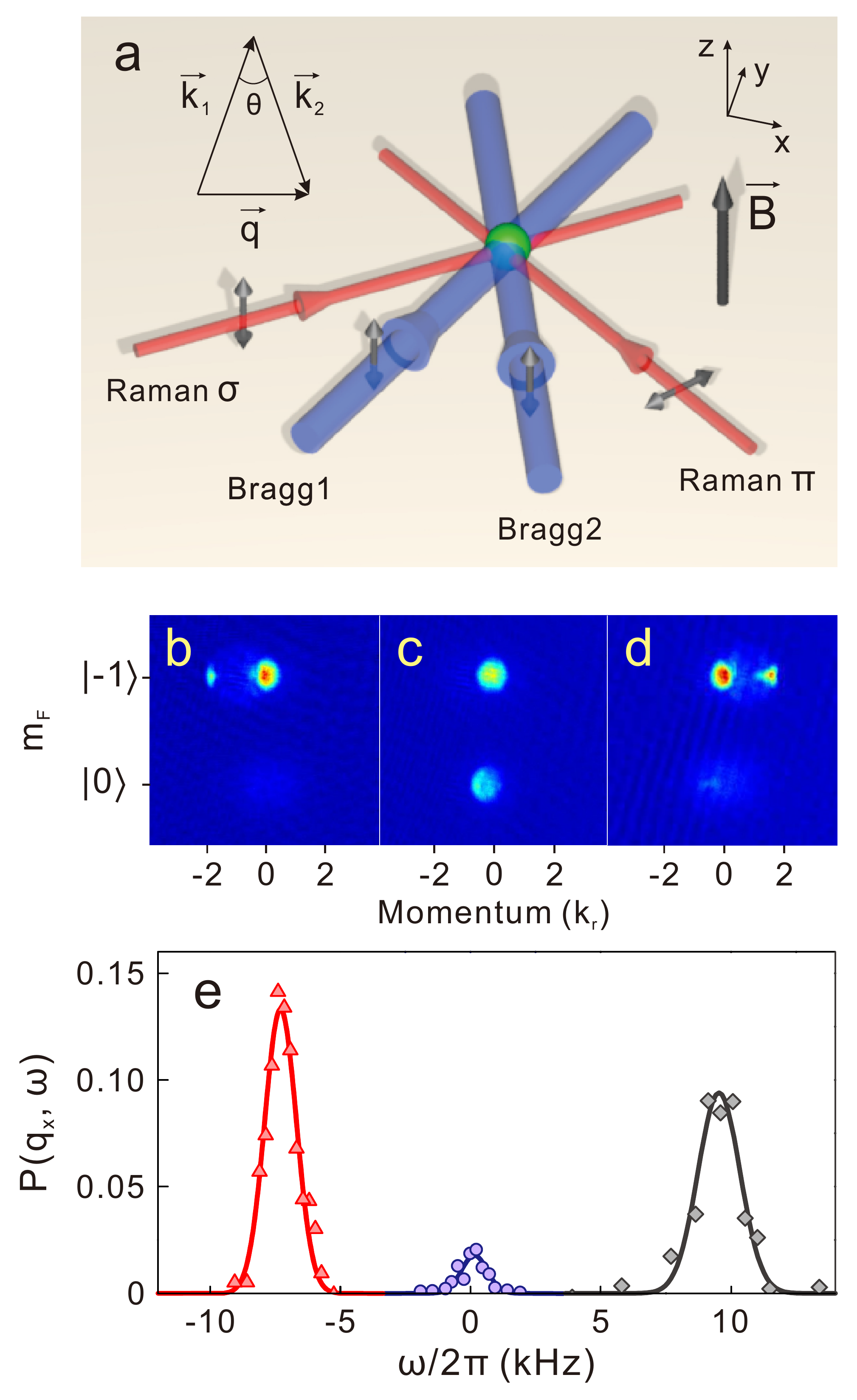}{}
\caption{Experimental setup and Bragg spectroscopy. ({\bf a}) 
In the $x$-$y$ plane, there are a pair of Raman lasers with relative angle $105^\circ$, and two Bragg beams 
separated by an angle $7^\circ\leq\theta\leq180^\circ$, which produce a momentum transfer in the $x$-direction with $0.16k_{\rm r}\leq|q_x|\leq2.60k_{\rm r}$.
A bias magnetic field in the $z$ direction generates the Zeeman splitting. 
({\bf b}-{\bf d}) Spin-resolved TOF images for $\Omega=2 E_{\rm r}$, with $q_x/k_{\rm r}=-1.77$ (b) and $1.77$ (c-d). 
({\bf e}) Excitation efficiency $P(q_x,\omega)$ for $\Omega=2E_{\rm r}$ with $q_x=-1.77k_{\rm r}$ (red triangles) 
and $q_x=1.77k_{\rm r}$ (purple circles and gray diamonds).
The curves correspond to the Gaussian fits. 
The two resonance peaks for $q_x=1.77k_{\rm r}$ are for the lower (purple circles) and the upper (gray diamonds) branch of excitation spectrums, respectively.
} \label{figure1}
\end{figure}

The experimental setup is sketched in Fig.~\ref{figure1}a, sharing much similarity to our previous work~\cite{Jinyi2012,Cong2014}.
A BEC of about $1.5\times10^5$ $^{87}$Rb atoms is prepared in a crossed dipole trap with frequency $\omega=2\pi\times\{45, 45, 55\}$Hz.
The two counter-propagating laser beams couple the three internal states of the F = 1 manifold, generated by a bias magnetic field. 
In addition, the BEC is illustrated by two Bragg beams with approximately parallel polarization, which are symmetric about the $y$ axis.

By a quadratic Zeeman shift $\varepsilon=4.53k_{\rm r}$ with recoil momentum $k_{\rm r}$, the $|m_\text{F}= 1\rangle$ state is effectively suppressed, 
and the system can be regarded as a spin-$1/2$ system. 
The single-particle Hamiltonian along the SO coupling direction (the $x$-direction) is given by ($\hbar=1$)
\begin{equation}\label{Hamilton}
H_0=\frac{(k_x-k_r\sigma_z)^2}{2m}+\frac{\delta}{2}\sigma_z+\frac{\Omega}{2}\sigma_x,
\end{equation}
where $m$ is the atom mass,  $\Omega$ is the Raman coupling strength, and $\delta$ is the two-photon detuning,
which is fine tuned to be $\delta=0$ in experiment. Symbols $\sigma_z$ and $\sigma_x$ represent the Pauli matrices, 
with $|m_\text{F}= -1\rangle$ for spin  $|\uparrow\rangle$ and  $|m_\text{F}= 0\rangle$ for spin  $|\downarrow\rangle$.
For each given $k_x$, Eq.~(\ref{Hamilton}) has two eigenstates 
with energy $\mathcal{E}_+ (k_x) >\mathcal{E}_- (k_x)$ for the upper ($+$) and the lower ($-$) branch of single-particle dispersion~\cite{zhengwei,Hui2014}, respectively.
The lower branch has two degenerate minima for $\Omega<4E_{\rm r}$ ($E_{\rm r}=k_{\rm r}^2/2m$), denoted by $\pm k_{\rm min}$, 
and has a single minima at $k_x=0$ for  $\Omega>4E_{\rm r}$.
With interatomic interactions of $^{87}$Rb atoms being taken into account, 
it has been shown~\cite{Spielman2011,Yun2012} that for $\Omega<0.2E_{\rm r}$, 
atoms condense in a superposition of two components with opposite momenta $\pm k_{\rm min}$, exhibiting the stripe order.
For $0.2E_{\rm r}<\Omega<4E_{\rm r}$, this system maintains the magnetized phase, where atoms condenses at  $k_{\rm min}$ or $-k_{\rm min}$. 
When $\Omega>4E_{\rm r}$,  the single-particle dispersion has only one single minimum at zero momentum, 
and the Bose gas hence exhibits no magnetization, i.e., the non-magnetized phase.

The excitation spectrum of the magnetized phase is measured through Bragg spectroscopy~\cite{Ketterle1999_1,Ketterle1999_2,Davidson2002,Vale2008}.
We first prepare the BEC at the spin state $|m_\text{F}= -1\rangle$, and adiabatically ramp up Raman coupling strength $\Omega$ to the desired value.
By this way, the condensate is at the minimum  $-k_{\rm min}$, where most of atoms are in the spin state $|m_{\rm F}=-1\rangle$. 
Then, we quickly switch on two Bragg lasers and hold them for $1\sim2$ms. The Bragg beams have wavelength 
about $\lambda_{\rm B}=780.24$nm and are detuned $6.8$GHz away from the resonance. 
The angle $\theta$ between the two lasers ~(Fig.~\ref{figure1}a)  determines the momentum transfer $q_x=2k_{\rm B}\sin(\theta/2)$ ($k_{\rm B}=2\pi/\lambda_{\rm B}$),
while the frequency difference $\omega$ is tuned to produce an excitation. 
The Bragg pulse kicks a small percent of atoms out of the condensate cloud.
The intensity of Bragg lasers is adjusted to excite at most $20\%$ atoms, such that the linear response theory applies~\cite{Stringari_book}.
Finally, with the Stern-Gerlach technique, we take spin-resolved time-of-fight (TOF) images  after 24 ms of free expansion.
Three examples for $\Omega=2 E_{\rm r}$  are shown in Fig.~\ref{figure1}b-d, which have $q_x/k_{\rm r}=-1.77$, 1.77, and 1.77, respectively.
The latter two have the same momentum transfer, but different frequency difference.
It is worth noting that the spins of atoms in  Fig.~\ref{figure1}c flip when being kicked out from the condensate by the Bragg pulse.
This is due to the lock of spin and momentum.

\begin{figure}
\centering
\includegraphics[width=0.95\columnwidth]{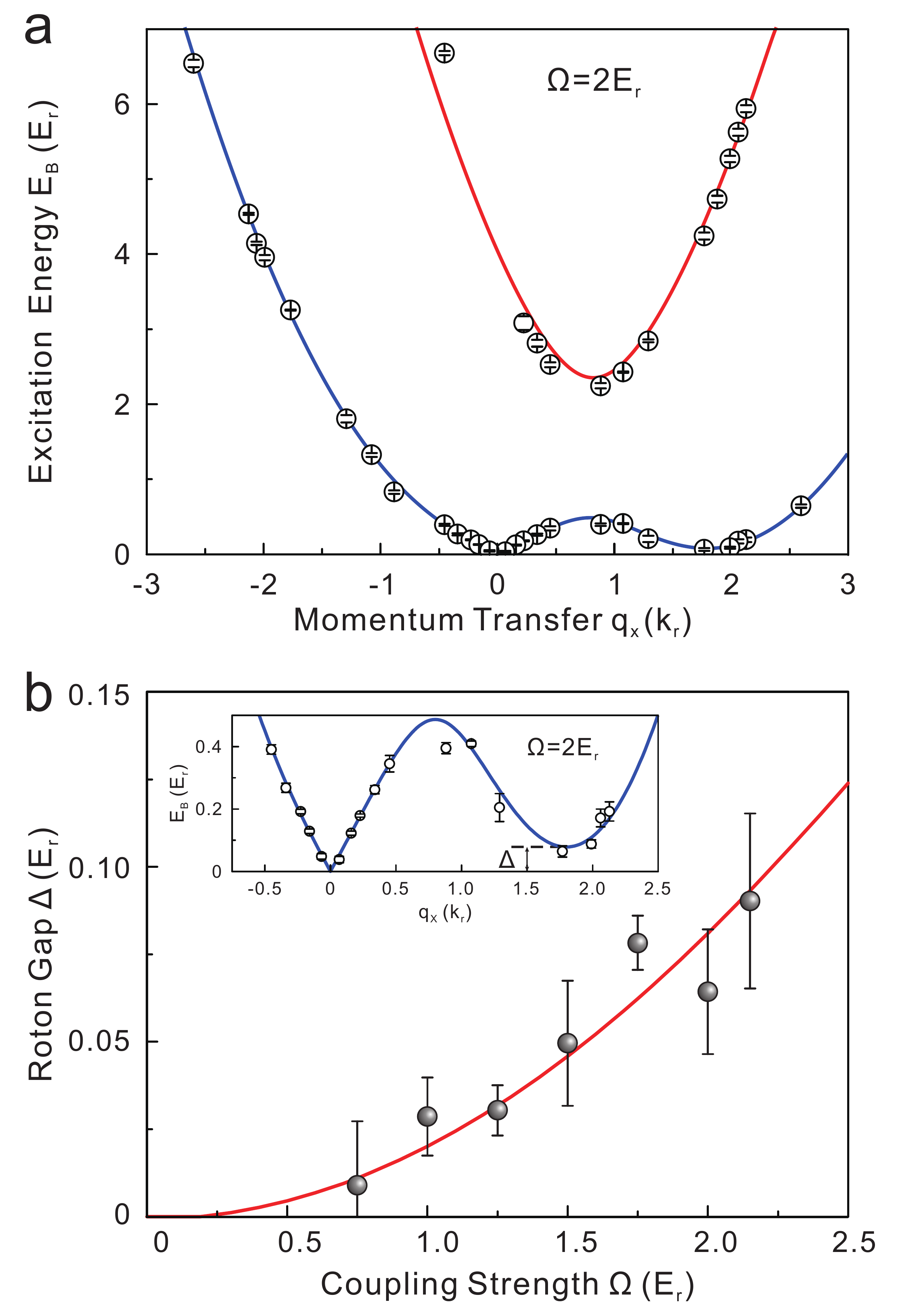}{}
\caption{Excitation spectrum and roton mode softening. ({\bf a}) Excitation spectrum for $\Omega=2E_{\rm r}$.  
The two branches of spectrums are clearly seen, and agree well with theoretical calculations (blue and red solid curves) based on a modified Bogoliubov theory.
 ({\bf b}) Softening of roton mode. 
 The inset of (b) shows the low-energy part of spectrum for  $\Omega=2E_{\rm r}$, clearly demonstrating a roton gap.
The measured roton gap $\Delta$ (circles) becomes smaller as $\Omega$ decreases,  and vanishes for small $\Omega$. 
} \label{figure2}
\end{figure}

For each TOF image, the atom numbers, $N_{\rm B}$ and $N_c$, of the Bragg and  the remaining condensate, are counted, 
and the ratio $\tilde{P}(q_x, \omega) \equiv N_{\rm B}/ (N_{\rm B}+N_c)$ is calculated.
According to the linear response theory~\cite{Stringari_book}, the excitation spectrum can be determined by the dynamic structure factor $S(q_x, \omega)$.
An evaluation based on Fermi's golden rule yields $\tilde{P}(q_x,\omega) \propto \Omega_B^2 S(q_x,\omega)$~\cite{Ketterle1999_2}, 
where $\Omega_B$ is the intensity of the Bragg lasers.
We define  excitation efficiency as $P(q_x,\omega)=\tilde{P}(q_x, \omega)/(\Omega_B/\Omega_{B0})^2 $, 
where  $\Omega_{B0}$ is chosen such that $ P(q_x,\omega)$ is a dimensionless quantity.
The excitation efficiency equals to the dynamic structure $P(q_x,\omega)=CS(q_x,\omega)$, apart from  an unknown constant $C$; 
For a given momentum transfer $q_x$, a broad range of frequency difference $\omega$ is scanned with Bragg laser intensity $\Omega_B$ being fixed at an appropriate value.
For $\Omega=2 E_{\rm r}$, Fig.~\ref{figure1}e shows the plot of the excitation efficiency $P(q_x,\omega)$ versus frequency difference $\omega$, 
for  $q_x/k_{\rm r}=-1.77 $ (red triangles)  and 1.77  (purple circles and gray diamonds).  
For simplicity and comparison purpose, $\Omega_{B0}$  is taken as the Bragg-laser intensity $\Omega_B$ for $q_x=-1.77 k_{\rm r}$.
It can be seen that there are two resonance peaks for momentum transfer $q_x/k_{\rm r}=1.77$, 
corresponding to the lower and the upper branch of the excitation spectrums, respectively.
The measured data are fitted by a Gaussian curve, and the peak frequency is used to identify the excitation energy.
The whole excitation spectrums for a fixed Raman coupling $\Omega$ are then constructed by varying momentum transfer $q_x$, 
see Fig.~\ref{figure2}a for $\Omega=2 E_{\rm r}$.

The excitation spectrum in the magnetized phase clearly shows a roton-type minimum at finite momentum around 
$q_x=2k_{\rm min}$ (see the inset of Fig.~\ref{figure2}b). We measure the roton gap $\Delta$, defined as the excitation energy 
at the roton minimum, and find it soften as Raman coupling strength decreases (Fig.~\ref{figure2}b). 
We calculate the roton gap based on a modified Bogoliubov theory~\cite{Stringari2012,zhengwei},  shown as the red solid curve in Fig.~\ref{figure2}b. 
The experimental data agree well with theoretical calculations.
As mentioned above, for $^{87}$Rb atoms, there is a phase transition near $\Omega_1 \approx 0.2 E_{\rm r}$ 
between the magnetized and the stripe phase, and accordingly, the roton gap is expected to vanish at $\Omega_1$.
Unfortunately, our experimental data are not sufficiently accurate to figure out the precise location of $\Omega_1$.
On the other hand, we do find that the roton-maxon structure disappears when $\Omega$ is tuned above a large enough value 
(about $3.4E_{\rm r}$ in our experiment), suggesting that the roton mode is a precursor of the stripe phase with periodic fringes.

The observed softening of roton gap can find its origin in Raman-dressed interaction.
In the presence of SO coupling,  interatomic interaction becomes anisotropic~\cite{Long2013}, 
and this anisotropy can be modified by tuning $\Omega$~\cite{Spielman_Science}.
This can be revealed by calculating the interaction energy for a condensate of different components. 
As shown in Ref.~\cite{Spielman2011}, for $^{87}$Rb atoms, we have interaction 
energy $E_{\rm I}\approx1/2\int d^3{\bf r}\{(c_0+c_2/2)[n_{\uparrow'}({\bf r})+n_{\downarrow'}({\bf r})]^2+
c_2/2[n_{\uparrow'}({\bf r})^2-n_{\downarrow'}({\bf r})^2]+(c_2+c_0\Omega^2/8E_{\rm r}^2)n_{\uparrow'}({\bf r})n_{\downarrow'}({\bf r})\}$, 
where the spin-independent
interaction $c_0=7.79\times10^{-12}$Hz$\cdot$cm$^3$, the spin-dependent interaction 
$c_2=-3.61\times10^{-14}$Hz$\cdot$cm$^3$, and $n_{\uparrow'}({\bf r})$ and $n_{\downarrow'}({\bf r})$ 
respectively represent the spatial density of the components at $-k_{\rm min}$ and $k_{\rm min}$.
This means  that the interaction energy for a condensate of two dressed components $\pm k_{\rm min}$ has 
additional energy terms compared to 
the energy for a single-component condensate at $k_{\rm min}$ or $-k_{\rm min}$.
Accordingly, one can give an estimation of the roton gap as $\Delta\approx c_0n(\Omega-\Omega_1)^2/16E_{\rm r}^2$ for $\Omega>\Omega_1$ with $n$ for the condensate density 
and $\Omega_1 \approx 0.2 E_{\rm r}$ marking the phase transition point between the stripe and the magnetized phase.

\begin{figure}
\centering
\includegraphics[width=0.95\columnwidth]{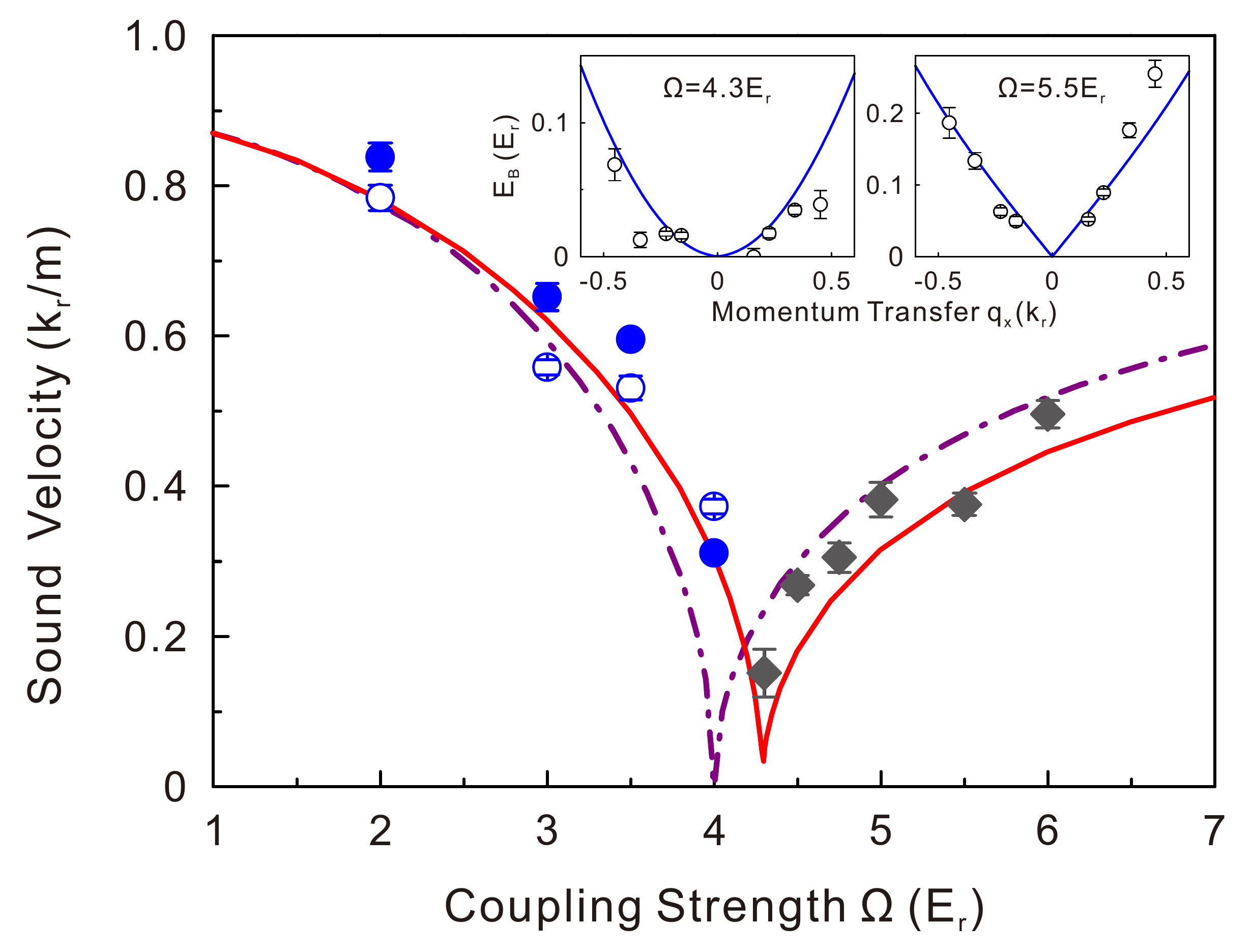}{}
\caption{ Softening of phonon mode. Sound velocities $c_1$ (solid circles) and $c_2$ (open circles) are both plotted for $\Omega<4.3E_{\rm r}$.
For $\Omega\geq4.3E_{\rm r}$, the sound velocity (diamonds) is taken as the average of the measurements in the negative and positive x-directions to minimize possible Doppler effect. 
Theoretical calculations for a spin-$1/2$ system (dot-dashed curve) predict a vanishing minimum at the phase transition point $\Omega=4E_{\rm r}$. 
A more practical consideration based on a spin-$1$ system with the effectively suppressed state $|m_{\rm F}=1\rangle$  (solid curve) shows that 
the sound velocity cannot touch zero, and the minimum is shifted to about $4.3E_{\rm r}$ for the set quadratic Zeeman shift $\epsilon=4.53E_{\rm r}$. 
The insets show the low-energy excitation spectrum for  $\Omega=4.3E_{\rm r}$ (left), where
the spectrum becomes parabola-like, and $\Omega=5.5E_{\rm r}$ (right), where linear mode recovers.
} \label{figure3}
\end{figure}

We also measure sound velocities both in the magnetized and the non-magnetized phase, and 
find a softening of the phonon mode near the phase transition between these two phases. 
In the magnetized phase, the excitation spectrum exhibits linear dispersions in the long wavelength limit, 
i.e. $E_{\rm B}(q_x)=-c_1q_x$ for $q_x<0$ and $E_{\rm B}(q_x)=c_2q_x$ for $q_x>0$; see the inset of Fig.~\ref{figure2}b.
Here $c_1$ ($c_2$) is the sound velocity in the negative (positive) $x$-direction. 
The measured velocities $c_1$ and $c_2$ are almost identical for a given $\Omega$ (Fig.~\ref{figure3}),
since the interaction difference in $^{87}$Rb atoms is very small~\cite{Stringari2012}.
The values of $c_1$ and $c_2$ decrease as $\Omega$ is enhanced, and reaches a minimum near the phase transition 
$\Omega_2$ that is slightly above $4E_{\rm r}$, as shown in Fig.~\ref{figure3}.
For $\Omega > \Omega_2$,  heating from Raman lasers makes it difficult to adiabatically load BEC into the minimum,  
and the condensate starts to oscillate in the trap during the Bragg pulse.
To minimize any effect of induced Doppler shift,
the sound velocity in the non-magnetized phase is taken as the average of the values for $q_x<0$ and $q_x>0$.
Nevertheless, it is still clear that the sound velocity increases with the coupling strength.

The non-monotonic behavior of the sound velocity as shown in Fig.~\ref{figure3} can be interpreted by the modification of 
single-particle dispersion. With the effective-mass approximation,
the sound velocity $c_s$ ($c_1=c_2=c_s$ is assumed) can be written as $c_s=\sqrt{gn/m^*}$~\cite{zhengwei}. 
Here the effective mass $m^{*}$ is given by  $m^{*}=m(1-\Omega^2/16E_{\rm r}^2)^{-1}$ for $\Omega<4E_r$
and $m^*=m(1-4E_{\rm r}/\Omega)^{-1}$ for $\Omega>4E_{\rm r}$. 
This shows that the vanishing of sound velocity originates in the divergency of the effective mass at $\Omega=4E_{\rm r}$, 
which marks the transition point between the magnetized and the non-magnetized phase. 
However, the Bose gas in our experiment is not a pure spin-$1/2$ system. 
Due to the influence of the suppressed state $|m_{\rm F}=1\rangle$, 
the value of sound velocity cannot drop to zero and the transition point is shifted to about $\Omega=4.3E_{\rm r}$ (Fig.~\ref{figure3}). 

Phonon mode softening indicates that at the transition point, the Bose gas should exhibit no superfluidity when  an impurity
moves inside with finite velocity in the SO coupling direction. It should be pointed out that
due to the absence of Galilean invariance~\cite{Jinyi2012} in SO coupled system, 
a moving SO-coupled Bose gas has different excitation spectrum from what we measure in Fig.~\ref{figure2}, and
softening of phonon mode is prevented in the comoving frame~\cite{zhengwei}.

\begin{figure}
\centering
\includegraphics[width=\columnwidth]{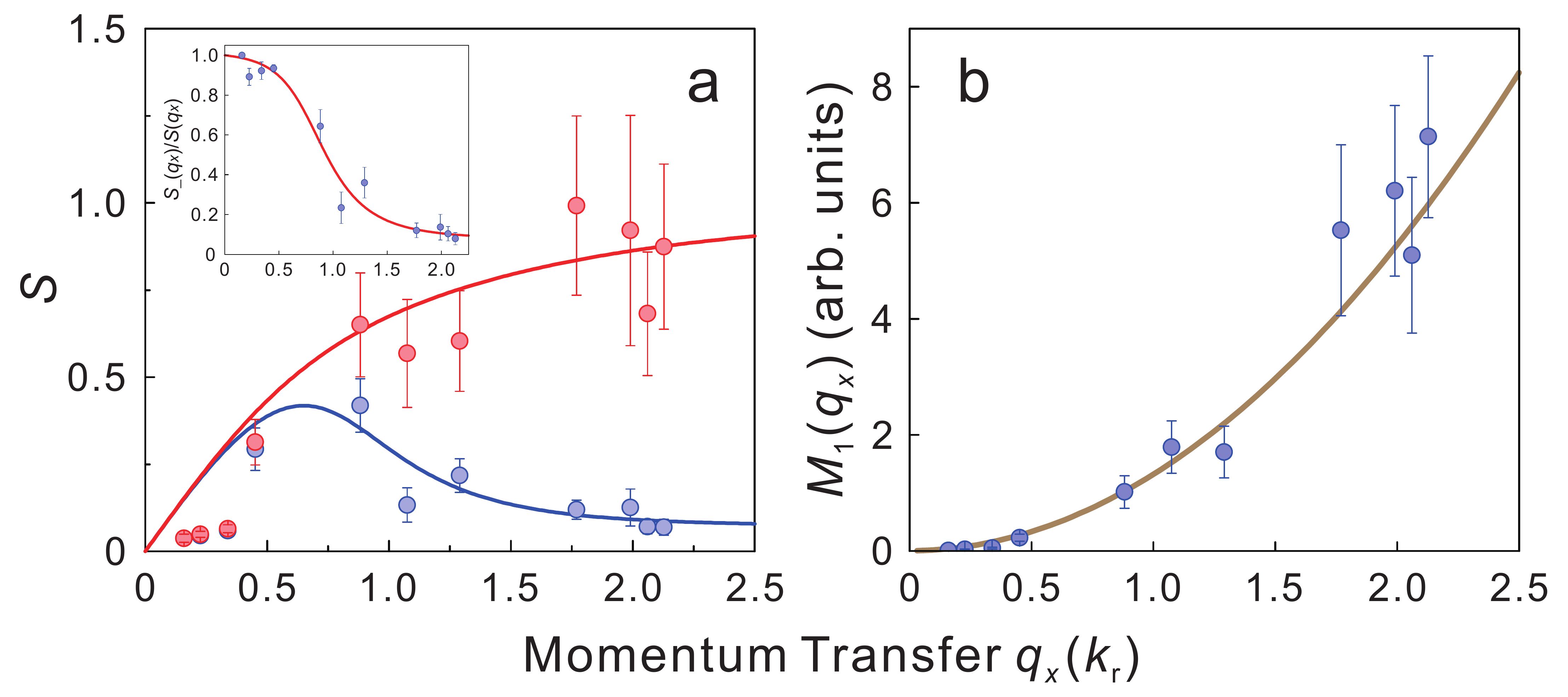}{}
\caption{Sum rules for $\Omega=2E_{\rm r}$. ({\bf a}) Static structure factor  $S(q_x)$ (red circles) and the contribution $S_{-}(q_x)$ (blue circles)
from the lower branch of excitation spectrum. The theoretical calculations based on location density approximations are shown as red and blue solid curves.
The relative contribution $S_{-}(q_x)/S(q_x)$ is shown in the inset.
({\bf b}) Energy-weighted moment $M_1(q_x)$. The good agreement between the experimental data (blue circles) 
demonstrates the validity of the $f$-sum rule.
} \label{figure4}
\end{figure}

Finally,  we examine sum rules in this unconventional system with SO coupling. 
Beside the excitation energy $E_B$ identified by the resonance peak position, 
the measured excitation efficiency $P(q_x, \omega)=C S(q_x, \omega)$ carries additional rich information.
This is  reflected by Fig.~\ref{figure1}e, where the heights of the three resonance peaks are significantly different.
The sum rules~\cite{Stringari_book} are about the behavior of moments of the dynamic structure, 
defined as $M_{p}({\bf q})\equiv\int\omega^pS({\bf q},\omega)d\omega = (1/C) \int\omega^p P({\bf q},\omega)d\omega $,
where $p \geq 0 $ is an integer. 
The zeroth-order moment for $p=0$  is proportional to the static structure factor as $M_0 (q_x) = N S(q_x)$,  where $N$ is the number of atoms. 
The famous $f$-sum rule is about the energy-weighted moment $M_1(\bf q)$, and states that $M_1(q_x)=N q_x^2/2m$, where $m$ is the mass of atom.
The validity of the $f$-sum rule in SO coupled system has been theoretically examined in Ref.~\cite{Stringari2012}.
In Fig.~\ref{figure4}a, we plot the static structure factor $S(q_x)$ for $\Omega=2E_{\rm r}$  as a function of $q_x$ (the red circles), 
where constant $C$ is chosen such that among the measured data, the maximum of $S(q_x)$ is equal to unity.
The blue circles in Fig.~\ref{figure4}a represent the contribution $S_{-}(q_x)$ from the lower branch of excitation spectrum.
The solid lines in Fig.~\ref{figure4}a are obtained from theoretical calculations based on local density approximation.
They agree with experimental data, except for those three points with very small momentum transfer $q_x$.
The relative contribution $S_{-}(q_x)/S(q_x)$ is shown in the inset of Fig.~\ref{figure4}a, 
which rapidly decreases as the momentum transfer becomes larger. 
The measured 1st moment $M_1(q_x)$ for $\Omega=2E_{\rm r}$ is plotted versus $q_x$ in Fig.~\ref{figure4}b. 
These experimental data can be well described by a quadratic curve, demonstrating the validity of the $f$-sum rule at least in the magnetized phase.

We have shown that despite interatomic interactions are weak and short-ranged, the SO coupled $^{87}$Rb condensate has
an excitation spectrum of roton-maxon character in the magnetized phase, which softens near the phase transition to the stripe phase. 
The sound velocities are also measured and a phonon-mode softening is observed. 
We mention that in condensed-matter physics and ultracold atomic physics, measurement of excitation spectrum is in itself of important role 
in revealing the properties of low-temperature phases~\cite{Griffin_book}.
The observed linear dispersion near $q_x=0$ is an important feature of superfluidity. 
Further, the measured roton-maxon structure of excitation spectrum, its disappearance for large $\Omega$, 
and the softening of the roton gap, strongly support the predicted ground-state phase diagram for of the SO coupled Bose gas of $^{87}$Rb atoms.


We acknowledge insightful discussions with H. Zhai, C. Chin, S. Stringari, Y. Li, and Z. Q. Yu. This work has been supported by the NNSF of China, the CAS, the National Fundamental Research Program (under Grant No. 2011CB921300).

$^{\star\star}$ These authors contribute equally to this work.
%



\end{document}